# Heuristic Algorithm using Internet of Things and Mobility for solving demographic issues in Financial Inclusion projects

**\*Subhamoy Chakraborti**
Magma Fincorp Limited, India
Email: chakrabortysubhamoy@gmail.com
**Sugata Sanyal**
Corporate Technology Office, Tata Consultancy Services, Mumbai, India
Email: sugata.sanyal@tcs.com
\*Corresponding author

## ABSTRACT
All over the world, the population in rural and semi-urban areas is a major concern for policy makers as they require technology enablement most to get elevated from their living conditions. Financial Inclusion projects are meant to work in this area as well. However the reach of financial institutions in those areas is much lesser than in urban areas in general. New policies and strategies are being made towards making banking services more accessible to the people in rural and semi-urban area. Technology, in particular Internet of Things (IoT) and Mobility can play a major role in this context as an enabler and multiplier. This paper discusses about an IoT and Mobility driven approach, which the Financial Institutions may consider while trying to work on financial inclusion projects.

**Keywords**
Financial Inclusion, Internet of Things, IoT, Mobile computing, Mobility, Enterprise Mobility, Financial Institution, NBFC, Microfinance

## 1. INTRODUCTION
Financial inclusion as a concept has been in practice for some time. In essence, Financial Inclusion entails delivering banking services to the low income and rural / semi urban people at an affordable cost [1]. Traditional Banks and Financial Institutions generally have presence in the urban areas. Also banks generally opt to choose customers with acceptable existing credit history. Thus a majority of rural and semi-urban customers, who do not have any formal credit history, remains outside the purview of Banks. A lot of Financial institutions operate in this scenario, who try to fill this outreach gap. These financial institutions, including Non-Banking Financial Corporations (NBFC) and Micro Finance Institutions (MFI), mostly operate in the rural and semi urban sector. Apart from the credit risk, these FIs also need to take decision fast while sourcing business from hinterlands of India.

Financial Inclusion, and thereby exclusion is a major issue all over the world, especially in the developing countries. India's central banking Institution, Reserve Bank of India, has mentioned the consequences of Financial Exclusions [1], which include increased unemployment, higher crime rate amongst others.

This paper attempts to discuss a technology framework which can help the FIs for reaching out to the customers and providing benefits faster. This paper shares overview of the technology strategy and a heuristic algorithm on image based decision making that can adopted by the Financial Institutions.

## 2. OVERVIEW
In Section 3, the paper discusses the technology model. In the subsections, detailed discussions covering various aspects of the technology stack, viz. Device choice, Mobile Application development framework, Application features and Network are covered. Also a heuristic algorithm on image handling is covered. Section 4 discusses about the overall Technology deployment architecture which stitches the various components discussed in Section 3. Section 5 concludes on the suggestions shared in the paper to implement a predictable IoT and Mobile enabled solution towards Financial Inclusion objective.

## 3. PROPOSED MODEL
There are many challenges in reaching out to the vast population who live outside the metro cities and towns. In this paper, we discuss about an Internet enabled device driven approach, through which the Financial Institutions can service the customer more predictably. This would increase the service quality as well as the reach of the FIs to the mass. To arrive at the proposed approach, the subsequent sections discuss about device options, options of application development framework, Application features, algorithms and network decisions.

### 3.1 DEVICE CHOICE
To serve the remote customers, most of the FI field executives used to depend on hand outs and printed forms. The Sales executives used to go to the customer and note down the requirements in their diaries and fill the requirement details in the loan application form. The field executive also used to collect the photocopy of the Know Your Customer (KYC) documents. The document collection process used to be an iterative process. Many a times the Sales executives needed to travel multiple times to the customer to collect all documents due to lack of knowledge of the Sales executive of the complete exhaustive list of required documents. Towards enabling the Sales executives via technology, initial innovation was towards using PDAs [3] in many FIs. Though they served a part of the requirement, availability of PDAs have gone down now drastically.

One alternate approach has been to use Point of Sale (POS) terminals [4]. It has been observed that the total cost for serving a customer reduces by use of the POS device [5]. However, these devices have a technology limitation due to the screen size. POS devices also operate on a comparatively low computing power. In view of the existing challenges, and the current emergence of Internet enabled devices, we have approached this problem with smart devices. The major change in Internet enabled device fabrication can help a lot in this context. This paper in particular discusses about an Android [6] smart device based approach.

Android have made devices affordable. Also the application development life cycle have got shorter thanks to the easy development process [7]. In view of this, FIs are better placed now to select Android based devices over POS devices. Android has grabbed a major market share [8] in the Mobile devices market. The market share continues to be over 80% of the devices [9] as per independent surveys. With the security aspects in mind [19] [20] [21] [22] of Internet of Things and the affordability index and the high market share of Android devices, the FIs are well positioned to make a choice of platform in favor of *Android* for embarking on Financial Inclusion projects.

### 3.2 APP DEVELOPMENT FRAMEWORK
The second step towards building the technology ecosystem is to select the Application development framework. There are many approaches in building am ubiquitous application. However based on surveys done [10], it has been found that CIOs and CTOs world over are choosing Native App development by a large margin over Web, Hybrid or any other mode of application development when it comes to building application impacting the business of the organization's success. While new approaches keep emerging, including Enterprise Suite, cross platform, cross compilation, HTML5 based application, FIs would be better positioned to go for Native Application approach in current context as this offers offline computing capability as well as better user experience, even if it may be a bit more complex to build Web or Hybrid approach [11], [12], [13], [14], [15].

To reach out to the rural and semi-urban customer base, it is absolutely necessary to build offline capability in the application in a secure way. In view of this requirement and the above observation, it is suggested that the FIs may choose *Native Android* Application development as its strategy while building Business to Employee applications.

### 3.3 APP FEATURES
A Financial Institution can design its Native Android Application majorly on three streams.

a) Capture the need details of the customer. This automates the information capturing process of the Sales executive on printed forms and diaries. This also removes a lot of rework where the back-office executives need to retype the details into digital form looking at the printed forms or removes the requirement for OCR scanner.
b) Collect images of the documents, especially Know Your Customer (KYC) documents. This step removes the multiple to-and-fro visits by the Sales executive to the customer place to collect various documents and submitting them to the back-office like Operations and Credit department for them to take a decision on loan approval.
c) Keep the Sales executive updated about the status of the loan application of the customer. This removes the requirement for the Sales executive to keep track of the loan applications and doing follow-up over phone with the back-office.

While point a) and c) mentioned above are transactional in nature, point b) requires an integration with the inbuilt back camera available with most ubiquitous devices and an algorithm to make the images ready for taking decisions upon. By taking photographs of the documents and sending the images electronically, it saves a lot of time of the

Sales executive and the customer. In any business process, *Motion* is identified as one of the causes of *wastage* as per Six Sigma methodology [16]. Through the proposed process of capturing images of documents, major reduction in Motion can be achieved, thus reducing wastage in the process and increasing turn-around-time. Earlier significant work has been done on improving the document capturing process [2]. The situation has improved due to the advent of smart phones with autofocus back cameras. It has been observed that a mobile device with a camera of 2MP resolution or higher can capture document images in daylight in legible format. Once the transactions and images are captured in the mobile device, it needs to be synched with the middleware application over mobile network.

### 3.4 IMAGE ENABLEMENT HEURISTIC ALGORITHM

Since the available Mobile devices for retail customers are fabricated keeping the consumers in mind, the images at times doesn't come as processed one. As a large amount of time gets spent in the *Motion* of the Sales folks collecting documents from customers in iterative manner, the image enablement of the documents in particular requires special attention. With the proposed algorithm in this paper, the image can become sharper and the size of the image can also come down.

This approach has dual impacts:
a) Better image helps faster decision by back-office users, thus helping the customer get the loan faster.
b) Smaller image size helps increase the probability of pushing the images over unpredictable or low bandwidth network faster

Suggested algorithm as listed below, can help enhance the images taken in the mobile device and help the Financial Institution take a faster decision on loan eligibility of the customer.

*Meta code*:

```
ImageProcPushback ()
{
    // Retrieve the image from application container within filesystem
    fetchImage (img)

    // Convert the color image to monochrome image. This would reduce the image size at first level
    convertColor (img, Monochrome)

    // Boundary detection would help to sharpen the boundaries of the image
    detectBoundary (img)
    sharpenBoundary (img)

    // Before streaming the images to the server at the backend, compress the images. A typical compression factor of 0.6 works for a 2MP camera. The compression factor needs to be decided based on the actual device and resolution used
    compressBitmap (img);

    // Queue images for pushing back to the server
    queueSend (img);

}
```

### 3.5 NETWORK
Connectivity has been improving in leaps and bounds over the last few years [18]. However last mile connectivity still remains a challenge outside the major cities and metros in most of the geographies. Though 3G connectivity is available in majority of the cities, the connectivity speed becomes unpredictable outside the metro and major cities. In view of this, choosing the right network provider is a key factor to make any such Mobile enablement initiative successful. In a country like India, the problem is compounded as no single private operator has 3G spectrum allocation covering the whole country.

In view of this, FIs are better placed to choose the network providers based on their areas of strength. Instead of choosing a single network provider to cover the whole country. This strategy can help in two ways:
a) Avoid dependency on a single network provider and thus a single point of failure for the entire business across the geography
b) Avoid local black holes of low network connectivity.

Hence based on field study, an FI is suggested to choose network providers having best connectivity in the specific state and locations.

### 4. PROPOSED DEPLOYMENT ARCHITECTURE

In this section we connect the separate pieces discussed in the above sections and depict a deployment architecture diagram which can be adopted as a blueprint.

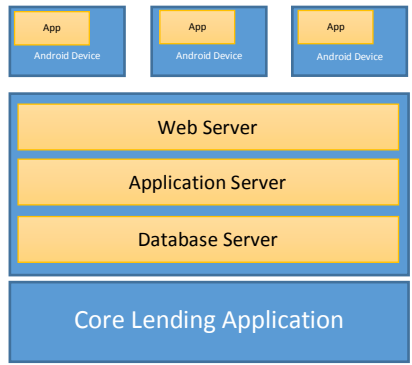

Fig 1: Mobile Sales Automation Product – Deployment Architecture

A typical ubiquitous technology stack can be represented as in Fig [1]. This is a suggestive deployment architecture. Android mobile devices can be used as the end user computing device. A native Mobile application built on Android stack can be installed on these devices, which can perform the features outlined in Section 3.3 App Features. The mobile application communicates with the middleware over GPRS/3G network. The middleware stack can be designed in 3 tier approach having Web Server, Application Server and Database Server. This stack in turn can connect to the Core Lending Application. This type of architecture helps reduce load at the mission critical transactional Core Lending Application. Also the sync engine built at Middleware abstracts out the devices from the Core Application. In all circumstances, it is not suggested to have the Mobile devices directly connect to the Core application.

## 5. CONCLUSION

Technology is a big enabler in increasing the reach of the Financial Institutions to the rural mass. With the increased availability of Mobile devices, attempts are being made to increase the reach of the Financial Institutions towards the Indian mass in a bigger scale. This technology enablement would help the FIs attain the Financial Inclusion objectives more fruitfully. To make these initiatives successful, problems needed to be solved in the Indian context, which are typical to this geography. Unless these challenges are tackled carefully, the technology solutions would not be able to serve the basic purpose. This paper shares some suggestions and highlights certain specific nature of Indian problem, which if adopted properly can lead to a predictable solution towards Financial Inclusion projects.

## Biographies and Photographs

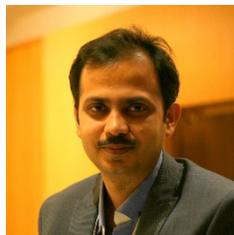

Subhamoy Chakraborti is working as General Manager at Magma Fincorp Limited, India. He has 12 years of experience in Technology Management. He has worked with Oracle India on ERP product development. He has also worked for various Tier 1 OEMs like Motorola, Toshiba, Fujitsu, Microsoft, Cisco, INQ in delivering phone programs while working for Product Engineering Services group in Wipro Technologies. He has keen interest in Data Science, Mobility and Cloud. His current focus includes building Enterprise Mobility ecosystem, Data driven decision making and Cloud Infrastructure. He has been working on defining the IT strategy and roadmap for his current organization with the CIO. Subhamoy participates in various campus focused programs and frequently contributes in CIO facing journals.

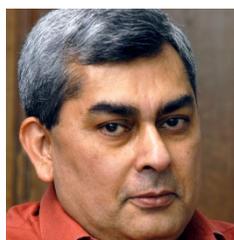

Sugata Sanyal is presently acting as a Research Advisor to the Corporate Technology Office, Tata Consultancy Services, India. He was with the Tata Institute of Fundamental Research till July, 2012. Prof. Sanyal is a: Distinguished Scientific Consultant to the International Research Group: Study of Intelligence of Biological and Artificial Complex System, Bucharest, Romania; Member, "Brain Trust," an advisory group to faculty members at the School of Computing and Informatics, University of Louisiana at Lafayette's Ray P. Authement College of Sciences, USA; an honorary professor in IIT Guwahati and Member, Senate, Indian Institute of Guwahati, India. Prof. Sanyal has published many research papers in International Journals and in International Conferences worldwide: topics ranging from network security to intrusion detection system and more.